\documentclass[4paper]{aipproc}

\layoutstyle{6x9}
  
\begin{document}

\title 
{A meson exchange model for the $YN$ interaction}


\author{J. Haidenbauer}{
  address={Forschungszentrum J\"ulich, IKP, D-52425 J\"ulich, Germany}
}

\iftrue
\author{W. Melnitchouk}{
  address={Jefferson Lab,
	12000 Jefferson Avenue, Newport News, VA 23606, USA} 
}

\author{J. Speth}{
  address={Forschungszentrum J\"ulich, IKP, D-52425 J\"ulich, Germany}
}
\fi

\copyrightyear  {2001}

\begin{abstract}
We present a new model for the hyperon-nucleon ($\Lambda N$, $\Sigma N$)
interaction, derived within the meson exchange framework.
The model incorporates the standard one boson exchange contributions of
the lowest pseudoscalar and vector meson multiplets with coupling
constants fixed by SU(6) symmetry relations.
In addition --- as the main feature of the new model --- the exchange of
two correlated pions or kaons, both in the scalar-isoscalar ($\sigma$)
and vector-isovector ($\rho$) channels, is included.
\end{abstract}

\date{\today}

\maketitle

\section{Introduction}

The hyperon-nucleon ($YN$) interaction is an ideal testing ground for
studying the importance of SU(3) flavor symmetry breaking in hadronic
systems.
Existing meson exchange models of the $YN$ force usually assume SU(3)
flavor symmetry for the hadronic coupling constants, and in some cases
\cite{Holz,Reu} even the SU(6) symmetry of the quark model.
The symmetry requirements provide relations between couplings of mesons of
a given multiplet to the baryon current, which greatly reduce the number
of free model parameters.
Specifically, coupling constants at the strange vertices are connected to
nucleon-nucleon-meson coupling constants, which in turn are constrained by
the wealth of empirical information on $NN$ scattering.
Essentially all $YN$ interaction models can reproduce the existing $YN$
scattering data, so that at present the assumption of SU(3) symmetry for
the coupling constants cannot be ruled out by experiment.

One should note, however, that the various models differ dramatically in
their treatment of the scalar-isoscalar meson sector, which describes the
baryon-baryon interaction at intermediate ranges.
For example, in the Nijmegen models \cite{NijIII,NijIV} this interaction
is generated by the exchange of a genuine scalar meson SU(3) nonet.
The T\"ubingen model \cite{Tueb}, on the other hand, which is essentially
a constituent quark model supplemented by $\pi$ and $\sigma$ exchange at
intermediate and short ranges, treats the $\sigma$ meson as an SU(3)
singlet.

In the $YN$ models of the J\"ulich group \cite{Holz,Reu} the $\sigma$
(with a mass of $\approx 550$~MeV) is viewed as arising from correlated
$\pi\pi$ exchange.
A rough estimate for the ratios of the $\sigma$-coupling strengths in the
various channels can then be obtained from the relevant pion couplings.
In practice, however, in the J\"ulich $YN$ models, which start from the
Bonn $NN$ potential, the coupling constants of the fictitious $\sigma$
meson at the strange vertices ($\Lambda\Lambda\sigma$,
$\Sigma\Sigma\sigma$) are free parameters --- a rather unsatisfactory
feature of the models.

These problems can be overcome by an explicit evaluation of correlated
$\pi\pi$ exchange in the various baryon-baryon channels.
A corresponding calculation was already performed for the $NN$ case in
Ref.~\cite{Kim}.
The starting point there was a field theoretic model for both the
$N\bar{N}\to\pi\pi$ Born amplitudes and the $\pi\pi$ and $K\bar{K}$
elastic scattering~\cite{Lohse}.
With the help of unitarity and dispersion relations, the amplitude for the
correlated $\pi\pi$ exchange in the $NN$ interaction was computed, showing
characteristic differences compared with the $\sigma$ and $\rho$ exchange
in the (full) Bonn potential.

In a recent study \cite{REUBER} the J\"ulich group presented a microscopic
derivation of correlated $\pi\pi$ exchange in various baryon-baryon
($BB'$) channels with strangeness $S=0, -1$ and $-2$.
The $K\bar{K}$ channel was treated on an equal footing with the $\pi\pi$
channel in order to reliably determine the influence of $K\bar{K}$
correlations in the relevant $t$-channels.
In this approach one can replace the phenomenological $\sigma$ and $\rho$
exchanges in the Bonn $NN$ \cite{MHE} and J\"ulich $YN$ \cite{Holz} models
by correlated processes, and eliminate undetermined parameters such as the
$BB'\sigma$ coupling constants.
As a first application of the full model \cite{PRELIM} for correlated
$\pi\pi$ and $K \bar K$ exchange, we present here new results for $YN$
cross sections for various $YN$ channels, and compare them with the
available data.

Alternative approaches to describing baryon-baryon interactions using
effective field theory, based on chiral power counting schemes, have
recently been applied to the $NN$ interaction.
However, at present a quantitative description of $NN$ scattering within
a consistent power counting scheme is still problematic \cite{MACH}.
Furthermore, it is not clear that such a description can be applied in
the strangeness sector, where the expansion parameter, $m_K/m_N$, may no
longer be small enough to allow an accurate low order truncation.
In addition, contact terms, which parameterize the intermediate and short
range interaction, do not fulfill any SU(3) relations, and cannot be
fixed by currently available $YN$ data.
At present, therefore, to obtain a quantitative description of $YN$
scattering data over a large energy range one is forced towards a more
traditional approach, such as that adopted here.

\section{Potential from Correlated $\pi\pi + K\bar K$ exchange}

Let us briefly describe the dynamical model \cite{Kim,REUBER} for
correlated two-pion and two-kaon exchange in the baryon-baryon
interaction, both in the scalar-isoscalar ($\sigma$) and vector-isovector
($\rho$) channels.
The contribution of correlated $\pi\pi$ and $K\bar K$ exchange is derived
from the amplitudes for the transition of a baryon-antibaryon state
($B\bar{B'}$) to a $\pi\pi$ or $K\bar K$ state in the pseudophysical
region by applying dispersion theory and unitarity.
For the $B\bar{B'} \rightarrow \pi\pi$, $K\bar{K}$ amplitudes a
microscopic model is constructed, which is based on the hadron exchange
picture.

The Born terms include contributions from baryon exchange as well as
$\rho$-pole diagrams (cf. Ref.~\cite{Janssen}).
The correlations between the two pseudoscalar mesons are taken into
account by means of a coupled channel ($\pi\pi$, $K\bar{K}$) model
\cite{Lohse,Janssen} generated from $s$- and $t$-channel meson exchange
Born terms. 
This model describes the empirical $\pi\pi$ phase shifts over a large
energy range from threshold up to 1.3 GeV.
The parameters of the $B\bar{B'} \rightarrow \pi\pi$, $K\bar{K}$ model,
which are interrelated through SU(3) symmetry, are determined by fitting
to the quasiempirical $N\bar{N'} \rightarrow \pi\pi$ amplitudes in the
pseudophysical region, $t \leq 4 m^2_\pi$ \cite{REUBER}, obtained by
analytic continuation of the empirical $\pi N$ and $\pi\pi$ data.

From the $B\overline{B'} \rightarrow \pi\pi $ helicity amplitudes one can
calculate the corresponding spectral functions (see Ref.~\cite{REUBER} for
details), which are then inserted into dispersion integrals to obtain the
(on-shell) baryon-baryon interaction in the $\sigma$ ($0^+$) and $\rho$
($1^-$) channels:
\begin{equation}
V^{(0^+,1^-)}_{B_1',B_2';B_1,B_2}(t) \propto \int_{4m^2_\pi}^\infty
dt' 
{\rho^{(0^+,1^-)}_{B_1',B_2';B_1,B_2}(t') \over t'-t}, \ \  t < 0 .
\end{equation}
Note that the spectral functions characterize both the strength and range
of the interaction.
For the exchange of an infinitely narrow meson the spectral function
becomes a $\delta$-function at the appropriate mass.

\section{Results and discussion}

As shown by Reuber et al.~\cite{REUBER}, the strength of the correlated
$\pi\pi$ and $K\bar{K}$ in the $\sigma$ channel exchange decreases as
the strangeness of the baryon-baryon channels becomes more negative.
For example, in the hyperon-nucleon systems ($\Lambda N$, $\Sigma N$) the
scalar-isoscalar part of the correlated exchanges is about a factor of 2
weaker than in the $NN$ channel, and, in particular, is also weaker than
the phenomenological $\sigma$ meson exchange used in the original J\"ulich
$YN$ model~\cite{Holz}.
Accordingly, we expect that the microscopic model with correlated $\pi\pi$
exchange will lead to a $YN$ interaction which is less attractive.

Besides replacing the conventional $\sigma$ and $\rho$ exchanges by
correlated $\pi\pi$ and $K\bar{K}$ exchange, there are in addition
several new ingredients in the present $YN$ model.
First of all, we now take into account contributions from $a_0(980)$
exchange.
The $a_0$ meson is present in the original Bonn $NN$ potential
\cite{MHE}, and for consistency should also be included in the $YN$
model.
Secondly, we consider the exchange of a strange scalar meson, the
$\kappa$, with mass $\sim 1000$~MeV.
Let us emphasize, however, that these particles are not viewed as being
members of a scalar meson SU(3) multiplet, but rather as representations
of strong meson-meson correlations in the scalar--isovector
($\pi\eta$--$K\bar K$) \cite{Janssen} and scalar--isospin-1/2 ($\pi K$)
channels, respectively.
In principle, their contributions can also be evaluated along the lines
of Ref.~\cite{REUBER}, however, for simplicity in the present model they
are effectively parameterized by one boson exchange diagrams with the
appropriate quantum numbers.
In any case, these phenomenological pieces are of rather short range,
and do not modify the long range part of the $YN$ interaction, which is
determined solely by SU(6) constraints (for the pseudoscalar and vector
mesons) and by correlated $\pi\pi$ and $K\bar K$ exchange.

In Fig.~1 we compare the integrated cross sections for the new $YN$
potential (solid curves) with the $YN \rightarrow Y'N$ scattering data
as a function of the laboratory momentum, $p_{lab}$.
The agreement between the predictions and the data \cite{DATA} is
clearly excellent in all channels.
Also shown are the predictions from the original J\"ulich $YN$ model~A
\cite{Holz} (dashed curves).
The main qualitative differences between the two models appear in the
$\Lambda p \rightarrow \Lambda p$ channel, for which the J\"ulich model
\cite{Holz} (with standard $\sigma$ and $\rho$ exchange) predicts a broad
shoulder at $p_{lab} \approx$ 350 MeV/c.
This structure, which is not supported by the available experimental
evidence, is due to a bound state in the $^1S_0$ partial wave of the
$\Sigma N$ channel.
It is not present in the new model.
The agreement in the other channels is equally good, if not better, for
the new model.
Further results and more details about the model can be found in
Ref.~\cite{we}.

\vfill \eject 

\begin{figure}[htb]
\includegraphics[height=.29\textheight]{cs1.epsi}
\includegraphics[height=.29\textheight]{cs2.epsi}
\includegraphics[height=.29\textheight]{cs3.epsi}
\end{figure}
\begin{figure}[htb]
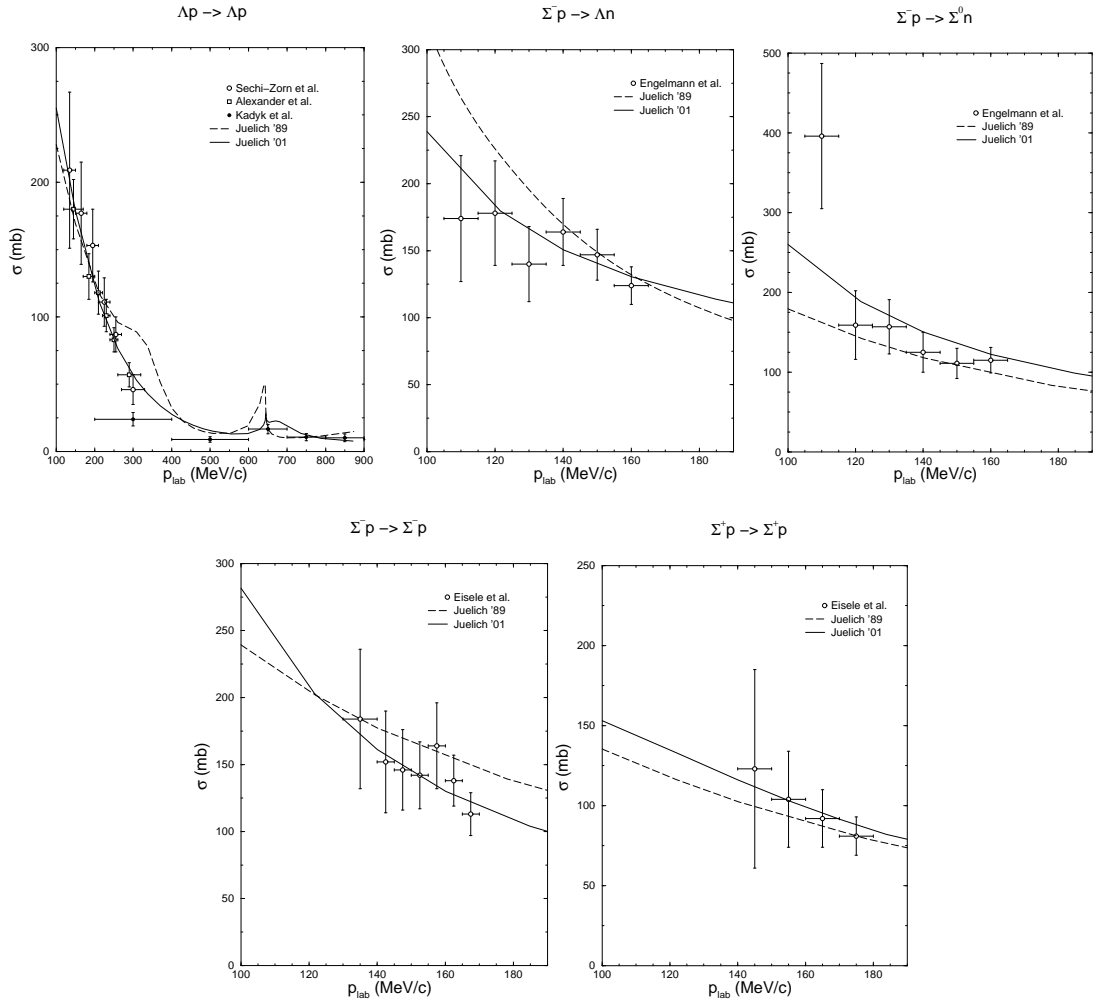

\includegraphics[height=.29\textheight]{cs4.epsi}
\includegraphics[height=.29\textheight]{cs5.epsi}
\caption{Total $YN$ scattering cross sections as a function of
	laboratory momentum, $p_{lab}$.
	The solid lines are results of the new $YN$ model, based on
	correlated $\pi\pi$ and $K\bar K$ exchange, while the dashed
	are results of the J\"ulich $YN$ model A \protect\cite{Holz}.
	The data are from Ref.\protect\cite{DATA}.}
\label{fig:cross}
\end{figure}

\def\Nucl{Nucl.\ }
\def\Phys{Phys.\ }
\def\Rev{Rev.\ }
\def\Lett{Lett.\ }
\def\PL{\Phys\Lett}
\def\PLB{\Phys\Lett B}
\def\NP{\Nucl\Phys}
\def\NPA{\Nucl\Phys A}
\def\NPB{\Nucl\Phys B}
\def\NPBS{\Nucl\Phys (Proc.\ Suppl.\ )B}
\def\PR{\Phys\Rev}
\def\PRL{\Phys\Rev\Lett}
\def\PRC{\Phys\Rev C}
\def\PRD{\Phys\Rev D}
\def\RMP{\Rev  Mod.\ \Phys}
\def\ZP{Z.\ \Phys}
\def\ZPA{Z.\ \Phys A}
\def\ZPC{Z.\ \Phys C}
\def\AOP{Ann.\ \Phys}
\def\PRep{\Phys Rep.\ }
\def\ANP{Adv.\ in \Nucl\Phys Vol.\ }
\def\PTP{Prog.\ Theor.\ \Phys}
\def\PTPS{Prog.\ Theor.\ \Phys Suppl.\ }
\def\PL{\Phys \Lett}
\def\JPF{J.\ Physique}
\def\FBSS{Few--Body Systems, Suppl.\ }
\def\IJMP{Int.\ J.\ Mod.\ \Phys A}
\def\NuCi{Nuovo Cimento~}

\begin{theacknowledgments}
This work is supported in part by the U.S. Department of Energy contract
\mbox{DE-AC05-84ER40150}, under which the Southeastern Universities
Research Association (SURA) operates the Thomas Jefferson National
Accelerator Facility (Jefferson Lab).
\end{theacknowledgments}

\end{document}